\documentclass[dvips]{article}
\usepackage{icrctc07}
\title{The H.E.S.S. survey of the inner Galactic plane}
\shorttitle{The H.E.S.S. survey of the inner Galactic plane}

\authors{S.~Hoppe$^{1}$,  for the H.E.S.S. collaboration}
\shortauthors{S.~Hoppe}
\afiliations{$^1$Max-Planck-Institut f\"ur Kernphysik, P.O. Box
  103980, D 69029 Heidelberg, Germany}

\email{stefan.hoppe@mpi-hd.mpg.de}

\abstract{The High Energy Stereoscopic System (H.E.S.S.), located in
  the Khomas Highlands of Namibia, is an array of four imaging
  atmospheric-Cherenkov telescopes designed to detect $\gamma$-rays in
  the very high energy (VHE; $>$ 100 GeV) domain. Its high sensitivity
  and large field-of-view (5$^{\mbox{\tiny o}}$) make it an ideal
  instrument to perform a survey within the Galactic plane for new VHE
  sources. Previous observations in 2004/2005 resulted in numerous
  detections of VHE gamma-ray emitters in the region l =
  330$^{\mbox{\tiny o}}$ - 30$^{\mbox{\tiny o}}$ Galactic
  longitude. Recently the survey was extended, covering the
  regions l = 280$^{\mbox{\tiny o}}$ - 330$^{\mbox{\tiny o}}$ and l =
  30$^{\mbox{\tiny o}}$ - 60$^{\mbox{\tiny o}}$, leading to the
  discovery of several previously unknown sources with high
  statistical significance. The current status of the survey will be
  presented.}

\begin{document}
\maketitle
\section{Introduction}
The majority of the newly discovered sources of very high energy (VHE;
$>$ 100 GeV) $\gamma$-rays are related to late phases of stellar
evolution, either directly to massive stars or to the compact objects
they form after their collapse. The possible associations include
pulsar wind nebulae (PWN) of high spin-down luminosity pulsars such as
G\,18.0$-$0.7 \cite{hess_j1825}, supernova remnants like
RX\,J1713.7$-$3946 \cite{RXJ1713}, and open star clusters like
Westerlund\,2 \cite{westerlund2}. As these objects cluster closely
along the Galactic plane, a survey of this region is an effective
approach to discover new sources and source classes of VHE
$\gamma$-ray emission.
\section{The H.E.S.S.  experiment and its Galactic plane survey}
\begin{figure}[t!]
\centering
\includegraphics[width=0.47\textwidth]{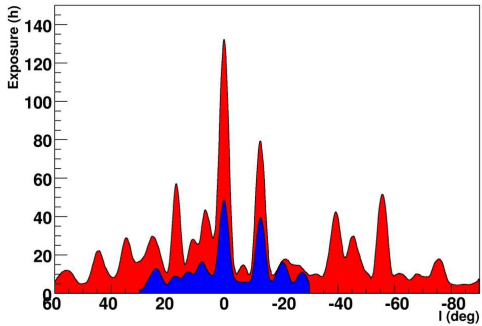}
\caption{Acceptance corrected livetime (equivalent time spent at an
  offset of 0.5$^{\circ}$) along the Galactic plane. All observations
  passing quality selection are considered, including survey-mode
  observations, re-observations of promising source candidates, and
  dedicated observations of known or expected VHE $\gamma$-ray
  sources. {\it Blue}: Observations taken in 2004/2005, published in
  \cite{hess_surveyI}. {\it Red}: Present status of data taking near
  the Galactic plane.
\label{fig:livetime}}
\end{figure}

The High Energy Stereoscopic System (H.E.S.S.) is an array of four
imaging atmospheric-Cherenkov telescopes located 1800~m above sea
level in the Khomas Highlands in Namibia \cite{hess_crab}. Each of the
telescopes is equiped with a camera comprising 960 photomultipliers
and a tesselated mirror with an area of 107\,m$^2$, resulting in a
comparatively large field-of-view of 5$^{\circ}$ in diameter. The H.E.S.S.
array can detect point sources at flux levels of about 1\% of the Crab
nebula flux near zenith with a statistical significance of 5$\sigma$
in 25 hours of observation. This high sensitivity and the large
field-of-view enable H.E.S.S. to survey large celestial areas -- such as
the Galactic plane -- within a reasonable time.\\
The H.E.S.S. Galactic plane survey began 2004 and has been a major
part of the observation program since. In the years 2004/2005 the
survey was conducted in the Galactic longitude band $\pm$ 30$^{\circ}$
around l = 0$^{\circ}$, covering most of the inner part of the
Galactic plane from the tangent of the Norma arm to the tangent of the
Scutum arm. Observations of 28 minutes duration each were taken at
pointings with a spacing of 0.7$^{\circ}$ in longitude in three strips
in Galactic latitude, covering an approximately 6$^{\circ}$ wide
region along the Galactic plane. 95\,h of data were taken in pure
survey mode. Promising source candidates were re-observed in dedicated
observations, comprising 30\,h of data. In addition, dedicated
observations in this region were taken on known or assumed VHE
$\gamma$-ray sources. The total amount of good quality data in this
region was 230 hours, Fig. \ref{fig:livetime} (blue). This first stage
of the H.E.S.S. Galactic plane survey resulted in the discovery of
eight previously unknown sources of VHE $\gamma$-rays at a
significance level greater than 6$\sigma$ after accounting for all
trials involved in the search (post-trials)
\cite{hess_survey_science}. Additionally, six likely sources were
found with significances above 4$\sigma$ \cite{hess_surveyI}.\\
In the years 2005-2007, the survey region was extended further along
the Galactic Plane. The scan region now covers $-$85$^{\circ}$ $<$ l $<$
60$^{\circ}$, $-$3$^{\circ}$ $<$ b $<$ 3$^{\circ}$, containing the
Carina-Sagittarius arm and part of the Perseus arm. In total,
$\sim$325\,h of data were taken in survey mode within this region,
together with 625\,h of pointed observations, which include
re-observations of source candidates and dedicated observations of
known or assumed VHE $\gamma$-ray emitters. Figure \ref{fig:livetime}
shows the present (red) and past (blue) exposure of the H.E.S.S. Galactic
plane scan.
\section{New sources of VHE $\gamma$-rays}
In the continuation of the H.E.S.S. Galactic plane survey, $>$14 new VHE
$\gamma$-ray sources were discovered so far at statistical significances
larger than 5$\sigma$ post-trials. The possible associations range
from young pulsars such as PSR\,J1846$-$258 (Kes 75), over middle-aged
pulsars (PSR\,J1913+1011) to a source first discovered at TeV energies
by the Milagro collaboration (MGRO\,J1908+06). A non-negligible fraction
of the sources, however, have no obvious counterparts.
\subsection{PSR\,J1846$-$0258 and Kes\,75}
\begin{figure}[t!]
\centering
\includegraphics[width=0.45\textwidth]{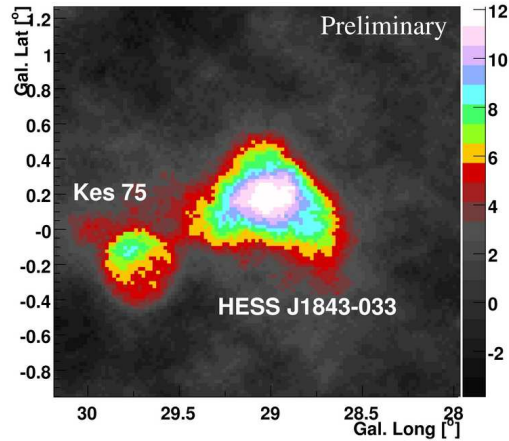}
\caption{$\gamma$-ray significance map of the region containing
  Kes\,75, obtained by counting $\gamma$-rays within 0.22$^{\circ}$
  from a given location. The integration radius is part of the
  standard survey analysis, chosen a-priori and therefore not adjusted
  to the individual source sizes. Significance values shown do not
  take the statistical trials involved in the survey into account.
  \label{fig:kes75}}
\end{figure}
The young shell-type supernova remnant (SNR) Kes\,75 is in many ways
similar to the well-studied Crab SNR. It contains the central pulsar
PSR\,J1846$-$0258, which powers an extended radio and X-ray core, and is
therefore another example of a centre-filled SNR, or plerion. Its
distance is estimated as $\sim$19~kpc \cite{kes75_extension}.
PSR\,J1846$-$0258 has a rotation period of 325~ms and a spin-down age
of 728~yrs \cite{atnf}, which apparently makes it the youngest
rotation-powered pulsar yet discovered \cite{kes75_age}. The
extensions of the core and the shell are 30'' and 3.5', respectively
\cite{kes75_extension}. Like from the Crab nebula, a point-like source
of VHE $\gamma$-ray emission is detected, coincident with the position
of Kes\,75, at a significance level of more than 8$\sigma$
post-trials. For details concerning the H.E.S.S. detection of this
object see \cite{proc_kes75}.
%
%
In the same field of view, an extended source HESS\,J1843$-$303 was
discovered with a statistical significance of more than 11$\sigma$
post-trials. In contrast to Kes\,75, no obvious counterpart for this
source was found yet, but a detailed archival search is still ongoing.
\subsection{HESS\,J1912+101}
Another possible example of VHE $\gamma$-ray emission from a PWN of a
high spin-down luminosity pulsar is HESS\,J1912+101, located at l =
44.4$^{\circ}$ and b = $-$0.1$^{\circ}$, detected at a post-trials
significance of more than 5$\sigma$. The corresponding pulsar,
PSR\,J1913+1011, is rather old, with a spin-down age of $t_{c} = 1.7
\times 10^{5}$~yrs, and nearby, at a distance of $\sim$4.5 kpc
\cite{atnf}. In contrast to PSR\,J1846$-$0258 mentioned earlier, no
PWN was detected during a dedicated Chandra observation of the
pulsar. The H.E.S.S. source is offset from the position of
PSR\,J1913+1011, which can be explained by either the proper motion of
the pulsar, or by an expansion of the PWN in an inhomogeous medium
\cite{pwn_asym}. The latter explanation seems plausible as clumpy
molecular material was found close to the pulsar position in
$^{13}$CO(J=1$\rightarrow$0) line measurements \cite{co13_survey}. The
scenario of an asymmetric PWN would make HESS\,J1912+101 similar to
HESS\,J1825$-$137 \cite{hess_j1825}.
\begin{figure}[t!]
\centering
\includegraphics[width=0.47\textwidth]{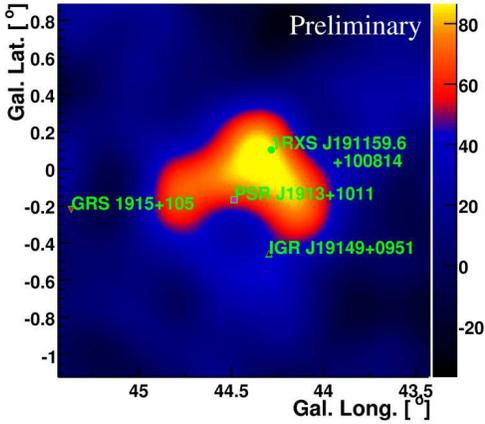}
\caption{Image of the VHE $\gamma$-ray excess from HESS\,J1912+101,
smoothed with a Gaussian profile of width 0.13$^{\circ}$. The
positions of the pulsar PSR\,J1913+1011, the ROSAT
source 1RXS\,J191159.6+100814, the INTEGRAL source
IGR\,J19149+0951 and the microquasar GRS\,1915+105 are marked. 
\label{fig:1912}}
\end{figure}
\subsection{MGRO\,J1908+06}
The Milagro collaboration, operating a ground-based air shower
detector near Los Alamos, announced the detection of several new
candidate emitters of TeV $\gamma$-rays in the Galactic plane
\cite{milagro}. Compared to the H.E.S.S. array, Milagro has a higher
energy threshold - the median energy of detected events is about 20
TeV - and a reduced angular resolution of about 1$^{\circ}$. The
Milagro coverage of the Galactic plane extends from about 30$^{\circ}$
longitude towards higher longitudes. Four sources are detected at
sufficient significance, the Crab Nebula and the new sources
MGRO\,2019+37, MGRO\,1908+06 and MGRO\,2031+41. Of these, only
MGRO\,1908+06, with a flux of 80\% of the Crab flux and a diameter of
up to 2.6$^{\circ}$, located at around 40$^{\circ}$ longitude is also
contained within the H.E.S.S. Galactic plane survey. Confirming the
Milagro result, a $\gamma$-ray source is detected with a significance
of more than 5$\sigma$ post-trials. The H.E.S.S. source is located at
l = 40.45$^{\circ}$ and b = $-$0.80$^{\circ}$, consistent with the
Milagro position of l = 40.4$^{\circ}$ and b = $-$1.0$^{\circ}$, with
an error radius on the Milagro position of 0.24$^{\circ}$. The rms
size of the H.E.S.S. source is about 0.2$^{\circ}$. For more details
on the H.E.S.S. result see \cite{proc_1908}.
\subsection{Unidentified sources}
A significant fraction of the recently discovered sources of VHE
$\gamma$-rays within the Galactic plane lack obvious counterparts. For
seven of these sources extensive archival searches in multi-wavelength
data and standard catalogues were performed to search for associated
objects in the radio, X-ray and GeV $\gamma$-ray domains.  While some
of them are partially coincident with known or unidentified X-ray
sources, none provide a clear counterpart which matches all of the
observed characteristics of the VHE emission. The lack of a
lower-energy counterpart challenges VHE emission scenarios, both
leptonic and hadronic. More details are given in a seperate contribution
to this conference \cite{proc_darksources} .
\section{Summary}
The H.E.S.S. Galactic plane survey, which started in the year 2004,
now reaches from $-$85$^{\circ}$ longitude to 60$^{\circ}$ longitude,
and covers an approximately 6$^{\circ}$ broad band around latitude b =
0$^{\circ}$.  In total, more than 950\,hours of data were taken in this
region, including survey mode observations, re-observations of source
candidates and dedicated observations of known or suspected
$\gamma$-ray sources. The first stage of the survey, covering the
inner 60$^{\circ}$ of the Galactic plane, has increased the number of
known VHE $\gamma$-sources within this region from three at the
beginning of 2004 to seventeen. Further follow-up observations within this
region and the extension of the survey along the Galactic
plane resulted in the discovery of even more additional VHE
$\gamma$-ray emitters. Most of them were presented during this
conference. Multi-wavelength follow-up observations and archival
searches have already begun, and will be crucial for understanding the
underlying processes at work in these astrophysical objects. 
\section{Acknowledgments}
The support of the Namibian authorities and of the University of Namibia
in facilitating the construction and operation of H.E.S.S. is gratefully
acknowledged, as is the support by the German Ministry for Education and
Research (BMBF), the Max Planck Society, the French Ministry for Research,
the CNRS-IN2P3 and the Astroparticle Interdisciplinary Programme of the
CNRS, the U.K. Science and Technology Facilities Council (STFC),
the IPNP of the Charles University, the Polish Ministry of Science and 
Higher Education, the South African Department of
Science and Technology and National Research Foundation, and by the
University of Namibia. We appreciate the excellent work of the technical
support staff in Berlin, Durham, Hamburg, Heidelberg, Palaiseau, Paris,
Saclay, and in Namibia in the construction and operation of the
equipment.\\

\bibliography{icrc0269}
\bibliographystyle{plain}
\end{document}